# Codes, Functions, and Causes: A Critique of Brette's Conceptual Analysis of Coding

David Barack[1], Andrew Jaegle[2]
[1]Columbia University, [2]DeepMind
[1]dlb2188@columbia.edu, [2]drewjaegle@google.com

## Abstract

In a recent article [1], Brette argues that coding as a concept is inappropriate for explanations of neurocognitive phenomena. Here, we argue that Brette's conceptual analysis mischaracterizes the structure of causal claims in coding and other forms of analysis-by-decomposition. We argue that analyses of this form are permissible, conceptually coherent, and offer essential tools for building and developing models of neurocognitive systems like the brain.

## 1. Coding and Causing

Brette argues that coding is an inappropriate concept for explanations of neurocognitive phenomena. Brette identifies three properties of coding: correspondence, representation, and causality. Brette grants correspondence but rejects both representation and causality for the neural code. While we disagree with his analyses of representation and causality, we limit our critique to the latter.

Brette's argument against causality focuses on two points. First, coding assumes that the parts of a cognitive system have separate functions. However, Brette claims that function cannot be attributed to the brain's parts. Second, coding implies linear causality for the brain. Brette argues instead that the brain features circular, coupled causality.

We argue that functions can be attributed to the parts of brains and, though brains are dynamical systems with circular causality, linear causality may still apply. We contend that the rejection of functions for parts of the brain constitutes a direct attack on the nature of explanation in cognitive neuroscience. Furthermore, the causality claim commits a category mistake, as the linear structure of the concept need not be mimicked by the causal structure of the brain. Finally, linear approximations are immensely successful in neuroscientific explanations.

## 2. Analysis by Decomposition

Brette first argues against the assignment of decoding and encoding functions to parts of the brain. Such assignment requires the analysis of behavior "...independently of the system in which the neurons are embedded" (p. 29). But such an analysis "… determines a neural code that... depends neither on the goals of the animals nor on the effect of spikes on the organism's actions" (p. 29-31). He concludes that the analysis of the brain by decomposition cannot proceed because "…function can be meaningfully ascribed to the organism as a system, but not... to the components of this system" (p. 31). Since such coding functions are defined independently of the organism's goals, Brette rejects the possibility of assigning coding functions to parts of the brain.

Brette's analysis relies on several misleading claims. First, some notions of function that do not rely on goals, such as causal role functions [2, 3], can be attributed to parts of organisms. Second, nothing about goals prevents function ascription to the organism's parts while permitting function ascription to the whole organism. As part of a larger system, the function of the part could share the goal of the organism. Indeed, this is typical for biology, where functions are often assigned to organs--such as the circulation of blood for the heart or cleaning the blood of toxins for the kidneys--even though the goals of these functions might be for the organism. Third, encoding and decoding can be specified with goals in mind. In particular, encoding and decoding can take into account variables relevant to the organism's biological fitness [4], and goal-oriented functions are defined and ascribed with respect to those fitness-relevant variables.

Furthermore, we contend that coding and other analysis-by-decomposition models are indispensable to explanations of brain function that integrate with psychology. These models break down psychological phenomena into their analytic subfunctions for explanation [cf. 5, 6, 7]. Decomposition requires that the psychological properties of subfunctions be reduced or removed when ascribing those subfunctions to component parts and proceeds recursively with finer decompositions with fewer psychological properties. At the lowest levels, functional descriptions are completely bare of psychology, yielding a reduction to neuroscience. Coding is a perfect example of such decomposition. The concept of coding implies encoding and decoding functions with reduced intentional implications, as those functions are grounded purely in probabilistic terms [8]. While message contents still need to be determined, coding analyses of systems like the brain can result in parts that carry weaker intentional properties.

## 3. Approximating Causal Structure

Brette next argues that the causality implied by coding does not apply to the brain. The neural code has linear causality (viz., input->encoder->decoder->output) whereas the brain possesses circular, coupled causality[1]. The brain is like a tent, where "...different elements are co-determined…. In addition to the coupling of neurons, the brain itself is coupled to its environment, i.e., there is circular and not linear causality" (p. 35). Linear causality refers to a temporally sequential, pairwise causally related sets of states whereas tent-like causality refers simultaneously occurring, jointly causally related sets of states. In short, the causal structure implied by the neural code fails to match the causal structure of the brain[2].

---

[1] Brette states two further causal structure problems that, while we disagree with both, considerations of length prevent addressing. First, the neural code is based on causally inefficacious properties (viz., spike rates and statistical quantities like averages) whereas only spatiotemporal particulars like individual spikes can be causally efficacious. Second, the properties identified by the neural code such as spike rates "abstract time away" (p. 35) whereas properties that set the brain's state occur at a time.

We first note that tent-like causality is consistent with linear causality. Linear encoding-decoding relationships between each pair of elements is consistent with an overall picture of a circular, coupled causal system. Indeed, Brette arguably commits a category mistake [9]: while the conceptual analysis of coding involves a linear structure [8], the implementations of encoding and decoding functions need not. This category mistake is further illustrated by Brette's claim that the temporal properties of coding structures are the "...discrete temporality of an algorithm, ...disconnected from physical time…. But… dynamical systems cannot in general be mapped to algorithmic descriptions" (p. 35). However, nothing about the coding metaphor entails discrete time, and one variable may encode another in the sense of carrying Shannon information without being part of an algorithm[3].

Brette's argument also ignores the utility of linearity and discrete time for analyzing complex systems. It is well known that continuous-time systems can be well approximated by discrete-time systems [10], and a host of equivalences exist between dynamical systems with circular causality and approximators with iterated linear causality [e.g.11, 12]. In practice, systems with tent-like causality can be iteratively or recurrently approximated by linear causal elements, such as by iteratively computing local relationships [e.g. 13, 14, 15]. Deep learning research uses discrete-time architectures to model many forms of continuous behavior as well (e.g. handwriting [16], speech synthesis [17], video prediction [18], robotic control [19], humanoid running [20]). Hence, complicated, ethologically relevant behavior can be analyzed with linear causality and discrete time, and this analysis is likely to remain a crucial part of building and designing models of this behavior [21].

---

[2] Note that the argument is overdetermined. The functional point and the causal point are independent, and, assuming that functions require causal powers, either would be sufficient from a causal point of view to reject the coding approach.

[3] Further, algorithms are in fact examples of a certain sort of discrete-time dynamical system and are run on systems with continuous dynamics at a lower level [22].